\documentclass{aastex63}

\usepackage{amsmath,amssymb}

\shorttitle{Dark matter admixed compact stars}
\shortauthors{Billy K. K. Lee et al.}
\graphicspath{{./}{figures/}}

\begin{document}

\title{Can the GW190814 secondary component be a bosonic dark matter admixed compact star?}

\author[0000-0001-6255-2773]{Billy K. K. Lee}
\affiliation{Department of Physics and Institute of Theoretical Physics, The Chinese University of Hong Kong, Shatin, N.T., Hong Kong SAR, People's Republic of China}

\author[0000-0002-1971-0403]{Ming-chung Chu}
\affiliation{Department of Physics and Institute of Theoretical Physics, The Chinese University of Hong Kong, Shatin, N.T., Hong Kong SAR, People's Republic of China}

\author[0000-0002-4638-5044]{Lap-Ming Lin}
\affiliation{Department of Physics and Institute of Theoretical Physics, The Chinese University of Hong Kong, Shatin, N.T., Hong Kong SAR, People's Republic of China}



\begin{abstract}

We investigate whether the recently observed 2.6 $M_\odot$ compact object in the gravitational-wave event GW190814 can be a bosonic dark matter admixed compact star. By considering the three constraints in mass, radius and stability of such an object, we find that if the dark matter is made of QCD axions, their particle mass $m$ is constrained to a range that has already been ruled out by the 
independent constraint imposed by the
stellar-mass black hole superradiance process. The 2.6 $M_\odot$ object can still be a neutron star admixed with at least 2.0 $M_\odot$ of dark matter made of axion-like particles (or even a pure axion-like particle star) if $2 \times 10^{-11}$ eV $\leq m \leq 2.4 \times 10^{-11}$ eV ($2.9 \times 10^{-11}$ eV $\leq m \leq 3.2 \times 10^{-11}$ eV) and with decay constant $f \geq 8 \times 10^{17}$ GeV.

\end{abstract}



\section{Introduction} 
\label{sec:intro}

Since the early indications of the existence of dark matter (DM) from observations of the galaxy rotation curves \citep{Rubin:1970zza} and galactic dynamics in clusters \citep{Zwicky:1933gu}, numerous observations including gravitational lensing \citep{Massey:2010hh} and cosmic microwave background \citep{Ade:2015xua} have confirmed the existence of this gravitating matter that interact weakly with normal matter (NM), which accounts for about 27\% of the total energy density in the universe today.

However, not many properties of DM are known. We have no clue about the spin-statistics, mass and interactions of the DM particles. A popular DM candidate is axion, a scalar bosonic particle originated from the study of quantum chromodynamics (QCD) \citep{Duffy:2009ig}. QCD axions, if exist, not only resolve the strong CP problem, but also can be a natural solution to the DM mystery. Axions can be produced in the early universe through various mechanisms such as thermal production \citep{Salvio:2013iaa}, misalignment mechanism \citep{Arias:2012az}, and from topological defects \citep{Kawasaki:2014sqa}. Even if QCD axions may not explain the current abundance of DM through the misalignment mechanism, axions can still be a part of DM. Other scalar bosons including axion-like particles (ALPs), motivated by string theory \citep{Arvanitaki:2009fg}, and repulsive self-interacting bosons, are also potential candidates of DM particles. Direct searches for these bosonic particles aim at exploiting their interactions with NM, such as a recent search for axions in the Xenon1T experiment \citep{Aprile:2020tmw}. However, no conclusive evidence has been found yet. Indirect searches for these bosonic particles have also been carried out, making use of the fact that DM are abundant in large scale. This includes searches for the self-annihilation signals of axions \citep{Zhu:2020tht} and bosonic DM effects on stellar objects \citep{Li:2012qf}. We focus only on the axions and ALPs in this paper.

A wide mass range of axions and ALPs has been considered in the literature.
For instance, ultralight axions with mass $m \sim 10^{-22}$ eV, also known as fuzzy DM, can solve
the core-cusp problem by forming DM halos with solitonic cores \citep{Hu:2000ke}. 
ALPs of $m \sim \mu$eV are also targets of laboratory-based experiments \citep{Braine:2019fqb, Anastassopoulos:2017ftl}.
QCD axions with $m > 1$ eV have been ruled out by different cosmological studies \citep{Fuentes-Martin:2019bue}, while $m \sim 10^{-11}$ eV that we are considering in this paper poses a great challenge for direct observations, which look for the axion-photon conversion signals that depend on the axion-photon coulping constant $g_{a\gamma \gamma}$. For $m \sim 2 \times 10^{-11}$ eV, the predicted coupling constant $g_{a\gamma \gamma} \sim 2.8 \times 10^{-21}$ GeV$^{-1}$ \citep{Beringer:1900zz} is far beyond the sensitivity of current experiments. It is noted that the ADMX experiment can constrain $g_{a\gamma \gamma} > 10^{-15}$ GeV$^{-1}$ \citep{Braine:2019fqb}.

Scalar bosons with Compton wavelengths comparable to the size of a rotating black hole would
lead to the superradiance process, which would extract angular momentum from the black hole and 
cause it to spin down rapidly \citep{Arvanitaki:2014wva, Zhu:2020tht,Hannuksela:2018izj,Baryakhtar:2020gao}.      
The existence of stellar-mass black holes with high spins has already ruled out scalar bosons of mass $6 \times 10^{-13}$ eV $\leq m \leq 2 \times 10^{-11}$ eV \citep{Arvanitaki:2014wva}.
QCD axions will also self-annihilate to produce continuous gravitational-wave signals. The lack of axion self-annihilation signals rules out $2 \times 10^{-13}$ eV $\leq m \leq 2.5 \times 10^{-12}$ eV \citep{Zhu:2020tht}.

Besides black hole superradiance, stellar-mass boson stars could also be an interesting probe of scalar bosonic particles of $m \sim 10^{-11}$ eV. Cold boson stars, first proposed in \citep{Kaup:1968zz, Ruffini:1969qy}, are formed by self-gravitating Bose-Einstein condensates. In such a star, all bosons share the same quantum state and form a macroscopic quantum state of astrophysical scale. Instead of the degeneracy pressure, the star supports itself by the Heisenberg uncertainty principle which cannot be described simply by an equation of state (EOS). Self-interacting boson stars with an interaction potential \(V(|\phi|) = \frac{1}{4} \lambda \left | \phi \right |^4 \) have also been considered in \citep{Colpi:1986ye}. It is shown that the maximum mass of the self-interacting boson stars can be substantially increased, comparing to the situation 
without interactions.
In the Newtonian limit, self-interacting boson stars are described by the Gross-Pitaevskii equation, which can be converted to the quantum Euler equations using the Madelung transformation. The self-interaction term acts like a polytropic EOS with polytropic index \(\gamma = 2\), and the quantum effect is accounted for by the quantum potential \(Q\) \citep{Chavanis:2011zi}. 
The general relativistic case that we consider in this work requires solving the scalar field equation coupled with the Einstein's equations \citep{Henriques:1989ar}.
Despite numerous theoretical studies, it is not clear whether boson stars exist in nature, though there are suggestions that neutron stars could be self-interacting boson stars \citep{Chavanis:2011cz}.

Besides pure boson stars, bosonic DM-admixed compact stars are also interesting, as they may form by accretion of DM onto compact stars \citep{DiGiovanni:2020frc}. Bosons can be excited in a neutron star by superradiance, similar to the case of a black hole \citep{Day:2019bbh}. There have also been discussions of the gravitational effects of self-interacting bosonic DM-admixed white dwarfs (WDs) and neutron stars (NSs) \citep{Li:2012qf}. However, simplifying assumptions such as neglecting the quantum potential or treating the DM as a concentrated core at the centre of the stars are made, which is not appropriate for QCD axions which have an attractive self-interaction. The formalism we use in this paper is suitable for all scalar bosons without the need of any simplification, but we will focus on QCD axions and ALPs.

The recent detection of the gravitational-wave event GW190814 \citep{Abbott:2020khf}
from the merger of a black hole of mass $\sim 23$ $M_\odot$ with a secondary component 
of mass $\sim 2.6$ $M_\odot$ is very interesting and unusual due to the extreme mass ratio of the system 
and the secondary component lying at the boundary of the so-called mass gap. 
Contrary to the first binary NS merger event GW170817 \citep{TheLIGOScientific:2017qsa}, no electromagnetic counterpart 
was identified for GW190814 and there was also no measurable matter effects in the gravitational-wave signal. From the mass measurement of the event GW190814 alone, the secondary can be either the lightest black hole or the heaviest NS ever observed. 
On the other hand, the signals from the GW170817 event have provided useful information about the nuclear matter EOS (e.g., \citep{Annala:2018,De:2018,Fattoyev:2018,Most:2018,Tews:2018}), and relatively soft EOSs are favoured. 
In particular, a maximum mass of $\sim 2.2 - 2.3$ $M_\odot$ for a nonrotating NS can be inferred from the GW170817 event \citep{Margalit_2017,Rezzolla_2018,Shibata_2019}. 
There thus exists a tension if one tries to explain the GW190814 secondary as a $\sim 2.6$ $M_\odot$ NS, which would require a much stiffer EOS than inferred from the GW170817 event. While different proposals have been suggested \citep{Abbott:2020khf,Most_2020,Essick_2020,Zhang_2020,Tsokaros_2020,Vattis_2020, Dexheimer_2021,Nathanail_2021}, 
it is still not certain whether the GW190814 secondary is a black hole, a neutron star, or a more exotic compact object, though it is likely a black hole according to recent studies that take into account multiple observation data and nuclear-physics constraints \citep{Fattoyev_2020,Tews_2021}. However, even if the GW190814 secondary is the lightest black hole ever observed, the first one in the mass gap, it still remains unclear how it can be formed in the first place (see \citep{Safarzadeh_2019,Safarzadeh_2020,Zevin_2020} for some proposals). 
In fact, whether the black-hole mass gap arises from physical origins or observational bias is also not clear at this point \citep{Ozel:2010su}.

In this paper, we study whether the $2.6$ $M_\odot$ compact object could be a bosonic DM-admixed NS (DANS) by constructing compact stars that are composed of bosonic DM particles (QCD axions
or ALPs) and normal nuclear matter. We choose the Akmal-Pandharipande-Ravenhall (APR) EOS \citep{Akmal:1998cf} to describe the nuclear matter. This EOS has a maximum nonrotating
NS mass of $2.2$ $M_\odot$ and is an example of relatively soft EOSs favored by the GW170817 event.
We employ the APR EOS to describe the nuclear matter as an illustration, but our conclusion in this work does not depend sensitively on the NM EOS, as long as it satisfies the 
condition that the resulting maximum NS mass is in the range $\sim 2.2 - 2.3$ $M_\odot$. Once the NM EOS model is fixed, the most important input parameters for our
computation of the structure of a DANS are then the boson mass, the self-coupling constant, and the amount of DM inside the star. We will survey the ranges of parameters that can potentially 
account for the $2.6$ $M_\odot$ GW190814 secondary. Interestingly, a large portion of the required parameter space for the QCD axions or ALPs has already been ruled out by black hole superradiance.

The plan of this paper is as follows.
In Section \ref{sec:method}, we present our method to find the equilibrium structures of DM-admixed compact stars. Our approach is general relativistic and the formalism we use does not assume any particular approximation limit for the scalar bosons. In Section \ref{sec:results}, we show the properties for these DM-admixed compact stars with different DM parameters, and only a limited DM parameter range could give rise to a DANS that can account for the GW190814 secondary. In Section \ref{sec:discussion}, we discuss the potential implications of our work, and a summary of our findings is given in Section \ref{sec:conclusion}.
We use units where $G=c=\hbar=1$ unless otherwise noted.

\section{Method} \label{sec:method}

Our starting point is the general self-interacting complex scalar field potential 
\begin{equation}
\label{eqn:1}
V(|\phi|^2) = \frac{1}{2} m^2 |\phi|^2 + \frac{1}{4}\lambda |\phi|^4,
\end{equation}
where $m$ is the mass of the scalar boson represented by the complex scalar field $\phi$, and $\lambda$ is the self-coupling constant. With this potential, the self-interacting scalar field Lagrangian density is given by \citep{Colpi:1986ye, Henriques:1989ar}
\begin{equation}
L = -\frac{1}{2}g^{\mu\nu}\partial_\mu \phi^* \partial_\nu \phi - \frac{1}{2}m^2 |\phi|^2 - \frac{1}{4} \lambda |\phi|^4.
\end{equation}
From the above Lagrangian density, we have the energy-momentum tensor for the bosonic DM
\begin{equation}
T_{\mu\nu, DM} = \frac{1}{2}(\partial_\mu \phi^* \partial_\nu \phi + \partial_\mu \phi \partial_\nu \phi^*) - \frac{1}{2} g_{\mu\nu}(g^{\rho\sigma} \partial_\rho \phi^* \partial_\sigma \phi + m^2 |\phi|^2 + \frac{1}{2}\lambda |\phi|^4).
\end{equation}

On the other hand, the energy-momentum tensor of the NM as a perfect fluid is given by
\begin{equation}
T_{\mu\nu, NM} = (\rho+p) u_\mu u_\nu + g_{\mu\nu} p,
\end{equation}
where \(\rho\) is the energy density, \(p\) is the pressure and \(u_\mu\) is the 4-velocity of the NM. As we consider the case where the DM and NM interact through gravity only, the energy-momentum tensors for both matters can be added up directly. In this point of view, the Einstein equations are
\begin{equation}
R_{\mu\nu} - \frac{1}{2} g_{\mu\nu} R = 8\pi (T_{\mu\nu, DM} + T_{\mu\nu, NM}).
\end{equation}
We consider spherically symmetric and static spacetime described by the 
metric
\begin{equation}
ds^2 = - B(r) dt^2 + A(r) dr^2 + r^2 d\Omega .
\end{equation}
We look for stationary solutions:
\begin{equation}
\phi(r,t) = \Phi(r) e^{-i \omega t},
\end{equation}
where $\Phi(r)$ is a real function. We also define the following variables:
\begin{equation}
\begin{split}
x = mr,\ \ \sigma = \sqrt{4\pi} \Phi,\ \  \Omega = \frac{\omega}{m}, \\
\Lambda = \frac{\lambda}{4\pi m^2}, \ \ \bar{\rho} = \frac{4\pi \rho}{m^2},\ \  \bar{p} = \frac{4\pi p}{m^2}.
\end{split}
\end{equation}
The resulting set of equations that governs a hydrostatic bosonic DM-admixed compact star is given by Eqs.~(18)-(21) of \citep{Henriques:1989ar}
\begin{equation}
\begin{split}
\frac{A'}{A^2 x} + \frac{1}{x^2}(1-\frac{1}{A}) = (\frac{\Omega^2}{B} + 1 )\sigma^2 + \frac{\Lambda}{2} \sigma^4 + \frac{\sigma'^2}{A} + 2\bar{\rho}, \\
\frac{B'}{ABx} - \frac{1}{x^2}(1-\frac{1}{A}) = (\frac{\Omega^2}{B} - 1 )\sigma^2 - \frac{\Lambda}{2} \sigma^4 + \frac{\sigma'^2}{A} + 2\bar{p}, \\
\sigma'' + (\frac{2}{x}+\frac{B'}{2B}-\frac{A'}{2A})\sigma' + A[(\frac{\Omega^2}{B}-1)\sigma - \Lambda \sigma^3] = 0, \\
\bar{p}' = -\frac{1}{2}\frac{B'}{B}(\bar{\rho}+\bar{p}),
\end{split}
\label{eqn:11}
\end{equation}
where the prime denotes \(d/dx\). The first two equations are obtained from the Einstein equations while the last two from energy-momentum conservation (\(\nabla_\nu T^{\mu\nu}_{NM} = \nabla_\nu T^{\mu\nu}_{DM} = 0\)). We look for a non-singular (\(\sigma'(0)=0\)), finite-mass (\(\sigma(\infty)=0\)) and asymptotically flat solution (\(B(\infty)=1\)). Two more boundary conditions are \(\sigma(0)=\sigma_0\) and \(\bar{\rho}(0)=\bar{\rho}_0\), corresponding to the central densities for the DM and NM. In addition, we require the solution to be node-less, which sets a constraint on the value of \(\Omega\). The two free parameters we put into the equations as initial conditions are \(\sigma_0\) and \(\bar{\rho}_0\).

In order to model real scalar field axions inside compact stars, we have $\phi = \phi^*$, and note that the well-known instanton potential expanded when $\phi << f$: \citep{Braaten_2019, Desjacques:2017fmf}
\begin{equation}
V(\phi) = (mf)^2 \left[1- \cos{ \left(\frac{\phi}{f} \right)} \right] = \frac{1}{2} m^2 \phi^2 + \frac{1}{4}\lambda \phi^4,
\end{equation}
where $f$ is the axion decay constant and $\lambda=-m^2/6f^2$, shares the same form as Eq. (\ref{eqn:1}). It should be remarked that this is not the QCD axion potential, though it agrees very well to the leading-order axion potential derived from the chiral effective theory when $\phi << f$, which we shall focus in this paper. For QCD axions, the following relation can be obtained from the chiral effective theory \citep{di_Cortona_2016}:
\begin{equation}
\label{eqn:3}
m_\pi f_\pi = \frac{1+z}{\sqrt{z}} mf,
\end{equation}
where $m_\pi = 135$ MeV is the pion mass, $f_\pi \sim 92$ MeV is the pion decay constant and $z = 0.56$ is the ratio of up and down quark masses \citep{Beringer:1900zz}. Hence, there is one less independent parameter for the QCD axions. However, for ALPs, $m$ and $f$ are two independent parameters. A negative $\lambda$ implies an attractive self-interaction. In the most general case of scalar bosons, $\lambda$ can also be positive representing a repulsive self-interaction.

The coupling of axions to photons ($L_{EM} = g_{a\gamma \gamma} \vec{E} \cdot \vec{B} \phi$, where $\vec{E}$ ($\vec{B}$) are electric (magnetic) fields) or neutrons are not significant in our scenario, as there are no visible axion cooling effects of neutron star for our interested parameter range \citep{Paul:2018msp}, and so we do not consider these terms in the Lagrangian density.

The total mass of the system $M_{total}$ can be extracted directly from the 
metric function by $A=(1-2 M_{total} / r)^{-1}$ at a point far away from the star so that the 
scalar field is vanishingly small. It can also be determined by 
using the Komar mass integral for stationary asymptotically flat spacetimes \citep{Wald:book}. 
In practice, the scalar field decreases sufficiently fast outside the star so that the Komar mass integral can be employed. An advantage to invoke the Komar mass integral is that the integral scales linearly with the total energy-momentum tensor $T_{\mu\nu}=T_{\mu\nu , DM} + T_{\mu\nu , NM}$. We can thus formally define the  
gravitational masses for DM and NM components of the star by the parts of the integral 
that involve $T_{\mu\nu, DM}$ and $T_{\mu\nu, NM}$, respectively. 
In particular, the gravitational mass of the NM component is given explicitly by 
$M_{NM} = 4\pi \int (\rho + 3p) \sqrt{AB} r^2 dr$. For convenience, instead of using the second part of the mass integral involving $T_{\mu\nu, DM}$, we calculate the gravitational mass of DM by $M_{DM}=M_{total}-M_{NM}$, where the total mass $M_{total}$ is extracted from the metric function 
as discussed above.

In the relativistic Thomas-Fermi (TF) limit, where the self-interaction is strong (\(\Lambda >> 1\)), we have the hydrostatic configuration described by the two-fluid Tolmann-Oppenheimer-Volkoff (TOV) equations \citep{Chavanis:2011cz, Mukhopadhyay:2015xhs, Oppenheimer:1939ne}

\begin{equation}
\begin{split}
\frac{dp_{NM}}{dr} = -\frac{GM_{enc}\rho_{NM}}{r^2}(1+\frac{p_{NM}}{\rho_{NM}})(1+4\pi r^3 \frac{p_{DM}+p_{NM}}{M_{enc}})(1-2\frac{M_{enc}}{r})^{-1}, \\
\frac{dp_{DM}}{dr} = -\frac{GM_{enc}\rho_{DM}}{r^2}(1+\frac{p_{DM}}{\rho_{DM}})(1+4\pi r^3 \frac{p_{DM}+p_{NM}}{M_{enc}})(1-2\frac{M_{enc}}{r})^{-1}, \\
\frac{dM_{enc, NM}}{dr} = 4\pi r^2 \rho_{NM}, \\
\frac{dM_{enc, DM}}{dr} = 4\pi r^2 \rho_{DM}, \\
M_{enc} = M_{enc, NM} + M_{enc, DM}, \\
p_{DM} = \frac{1}{36K}[(1+12K\rho_{DM})^{1/2}-1]^2, \\
K = \frac{\lambda}{4m^4} = \frac{\pi \Lambda}{m^2},
\end{split}
\label{eqn:12}
\end{equation}
where $M_{enc, NM}$, $M_{enc, DM}$, and $M_{enc}$ are the enclosed mass at $r$ of NM, DM, and total mass, respectively. As a result, $M_{total}$ ($M_{NM}$, $M_{DM}$) will be $M_{enc}$ ($M_{enc, NM}$, $M_{enc, DM}$) at radial infinity. This set of equations is solved under the boundary conditions \(\rho'_{NM}(0) = 0, \rho'_{DM}(0) = 0, M_{enc, NM}(0)=0, M_{enc, DM}(0)=0\), \(\rho_{NM}(0) = \rho_{0,NM}\) and \(\rho_{DM}(0) = \rho_{0,DM}\). The two free parameters are \(\rho_{0,NM}\) and \(\rho_{0,DM}\).

In the low-density limit, Eq. (\ref{eqn:12}) reduces to the Newtonian case, while in the relativistic limit, the EOS for the bosonic DM is described by a polytropic EOS with \(\gamma = 1\). We switch from Eq. (\ref{eqn:11}) to Eq. (\ref{eqn:12}) when $\Lambda$ is large enough such that the difference between $M_{total}$ in Eq. (\ref{eqn:11}) and Eq. (\ref{eqn:12}) is smaller than 1\%. 
Eq. (\ref{eqn:12}) has the advantages that it is computationally more efficient 
and the physics is more transparent. For instance, we can see directly from Eq. (\ref{eqn:12}) that in the TF limit the only dependent parameter is $m/\sqrt{\Lambda}$.

In the Newtonian description, there are three forces acting on the bosonic DM component of the DANS: the repulsive quantum pressure, which is a pure quantum effect, the pressure gradient force arising from the self-interaction, which can be attractive or repulsive, and the attractive gravitational force, from both the NM and DM. This description will be helpful in understanding the hydrostatics and hydrodynamics of the DANS. On the other hand, the NM component of the DANS has two forces balancing each other, the repulsive pressure gradient force and the attractive gravitational force.

Eq. (\ref{eqn:11}) and Eq. (\ref{eqn:12}) are closed by choosing an EOS for the NM. The APR EOS \citep{Akmal:1998cf, Schneider:2019vdm} is chosen for the high density region, while the SLy4 \citep{Douchin:2001sv} and Baym-Pethick-Sutherland  (BPS) EOSs \citep{Baym:1971pw} are chosen for the low density region in this paper. With the above EOSs, we have a full range of NM densities covering WDs to NSs, which will be needed in our calculations.

\section{Results} \label{sec:results}

As the maximum mass of the NM EOS is $\sim 2.2$  $M_\odot$, at least 0.4 $M_{\odot}$ of DM is admixed to the NS in order to account for the 2.6 $M_{\odot}$ compact object. It is known that the maximum mass of a pure boson star \(M_{max, DM}\) with attractive self interaction scales with $\lambda$ as \(M_{max, DM} \propto |\lambda |^{-1/2}  \) \citep{Chavanis:2011zi}, and $M_{max, DM} \propto m^{-2}$ for QCD axions. A rough estimation using $m \sim 2 $ $\mu$eV gives only $M_{max, DM} \approx 2 \times 10^{-10}$ $M_\odot$ for QCD axions. Unless the bosons have a large repulsive self-interaction, $m > \mu$eV is not preferred in our scenario. For QCD axion, the favourable range of \(m\) is $m \sim 10^{-11}$ eV.

\begin{figure}
\centering
\includegraphics[width=1.0\textwidth]{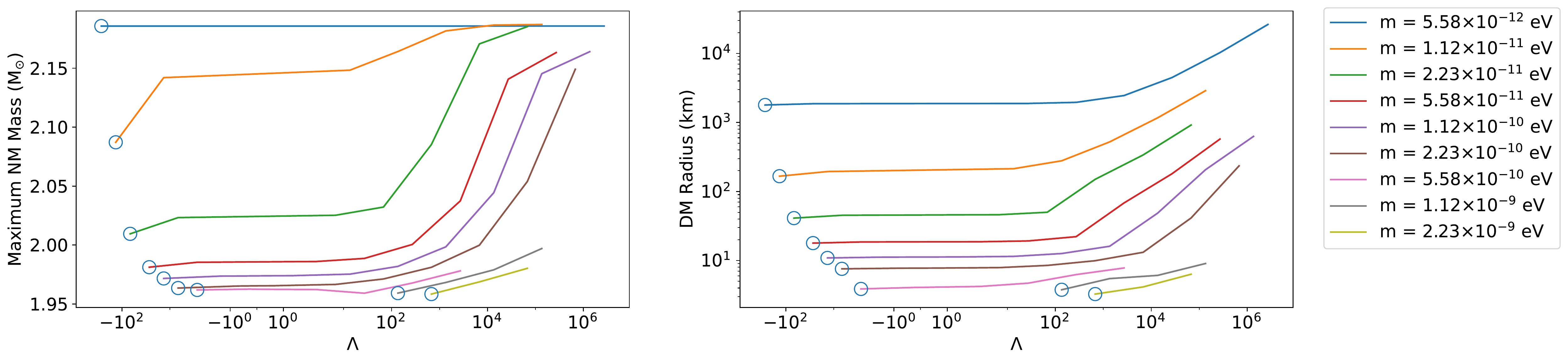}
\caption{Maximum NS mass \(M_{max,NS}\) (left) and the corresponding radius of DM \(R_{DM}\) (right) with 0.1 \(M_\odot\) of bosonic DM admixed, for different boson mass \(m\) and self-coupling constant \(\Lambda\).}
\label{fig:1}
\end{figure}

There are three parameters that affect the equilibrium structure of the DANSs: $m$, $\Lambda$ and $M_{DM}$. We first fix $M_{DM}$ and vary $m$ and $\Lambda$ to understand how they will affect the NM properties in a DANS. The left panel of Figure \ref{fig:1} shows the maximum NS mass $M_{max,NS}$ for different $m, \Lambda$ with $M_{DM} = $ 0.1 $M_\odot$. 

It is clear that when $m \leq 5.58 \times 10^{-12}$ eV, for any values of $\Lambda$, $M_{max,NS} = 2.19$ $M_\odot$ corresponding to the same $M_{max,NS}$ for the case of a NS with no DM admixed. The case with a larger $m$ will be different. There are two particular limits that are worth noting. One is the non-interacting limit ($\Lambda \sim 0$), where the only two forces acting on the DANS are the repulsive quantum pressure and the attractive gravitational force. In this limit, the only dependent parameter is $m$ for a fixed $M_{DM}$. The other limit is the TF limit ($\Lambda >> 1$), where the repulsive self-interaction is strong. The Newtonian description for this case is that the quantum pressure can be neglected and the only two forces acting on the star are the pressure gradient force arising from the repulsive self-interaction and the attractive gravitational force. As discussed previously, the only dependent parameter is 
$m / \sqrt{\Lambda}$ in this case.

In the non-interacting limit ($\Lambda \sim 0$), as $m$ increases, $M_{max,NS}$ decreases correspondingly. In the TF limit ($\Lambda >> 1$), $M_{max, NS}$ always decreases as $m/ \sqrt{\Lambda}$ increases. With $m = 1.12 \times 10^{-11}$ ($1.12 \times 10^{-10}$) eV in the non-interacting limit, $M_{max, NS}$ is decreased to 2.15 (1.97) $M_\odot$. Similarly, for $m/ \sqrt{\Lambda} = 3.4 \times 10^{-13}$ ($8.6 \times 10^{-13}$) eV in the TF limit, $M_{max, NS} = 2.14$ (2.05) $M_\odot$.

The reason for the decrease in $M_{max,NS}$ when \(m\) ($m / \sqrt{\Lambda}$) increases in the non-interacting (TF) limit is that the DM becomes more compact. The right panel of Figure \ref{fig:1} shows the radius of the DM, $R_{DM}$, at $M_{max,NS}$ for different values of $m$ and $\Lambda$. Since the density profile of the bosonic DM extends formally to infinity, we define $R_{DM}$ to be the radial location where $\rho_{DM}$ drops below $10^{-6}$ of $\rho_{0,DM}$. Even with a lower cutoff density for $R_{DM}$, the enclosed DM mass at $R_{DM}$ would not change by more than 1\%. We can infer from Figure \ref{fig:1} that with a smaller $R_{DM}$, the DM's gravitational effect will be stronger leading to a smaller \(M_{max,NS}\).

Note that when $m > 5.58 \times 10^{-10}$ eV, there is no longer a stable DANS for $\Lambda < 0$ as $M_{max, DM}$ reaches 0.1 $M_\odot$. For $\Lambda > 0$, as long as the repulsive self-interaction is strong enough, there can be a DANS with $M_{DM} = 0.1$ $M_\odot$ for $m > 5.58 \times 10^{-10}$ eV. The minimum $R_{DM}$ is achieved when $M_{DM} = M_{max, DM} = 0.1$ $M_\odot$, which results in the maximum compactness of the DM. This creates the maximum gravitational effect on the NM and results in the maximum decrease in $M_{max,NS}$. With $M_{DM} = $ 0.1 \(M_\odot\), the maximum decrease of $M_{max,NS}$ is from 2.19 $M_\odot$ to 1.96 $M_\odot$ with our EOS.

\begin{figure}
\centering
\includegraphics[width=0.8\textwidth]{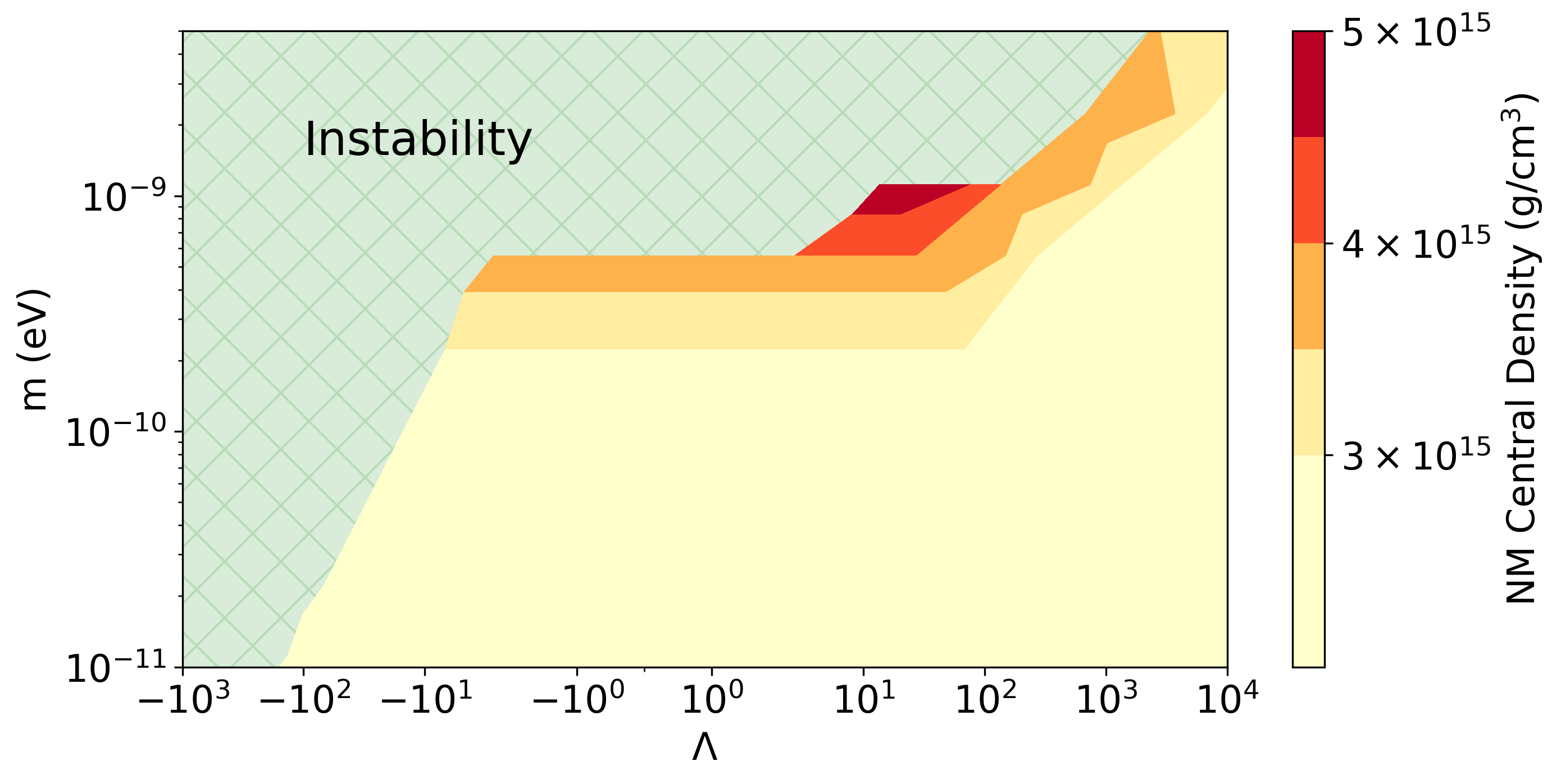}
\caption{NM central density \(\rho_{0,NM}\) at \(M_{max,NS}\) with 0.1 \(M_\odot\) of bosonic DM admixed, for different boson mass \(m\) and self-coupling constant \(\Lambda\). The green region marks the parameters where the bosonic DM will collapse into a black hole due to instability ($M_{max, DM} < 0.1$ $M_\odot$).}
\label{fig:2}
\end{figure}

Also, there exists a minimum value of $\Lambda$ for a given value of $m$ and $M_{DM} = 0.1$  $M_\odot$, which is indicated by the circle on the left end of each line in Figure \ref{fig:1}.
Below this minimum $\Lambda$, the attractive self-interaction together with the gravitational force are strong enough to overcome the repulsive quantum pressure, and the star collapses. This instability region can be more clearly seen in Figure \ref{fig:2}. 

In Figure \ref{fig:2}, we plot $\rho_{0,NM}$ at $M_{max, NS}$ in a DANS with $M_{DM} = $ 0.1 $M_\odot$ for different values of $m$ and $\Lambda$. We can see that $\rho_{0,NM}$ increases when $m$ is larger ($m > 2 \times 10^{-10}$ eV when $\Lambda \sim 0$). In the extremal case, $\rho_{0, NM}$ could be around 2 times higher than the maximum central density of an ordinary NS with no DM admixed ($\sim2.5 \times 10^{15}$ g/cm$^3$) with $m \sim 10^{-9}$ eV and $\Lambda \sim 10$.

Not all values of $m$ and $\Lambda$ can sustain stable DANSs. The green region in Figure \ref{fig:2} marks the parameter values that the DM will collapse into a black hole ($M_{max, DM} < 0.1$ $M_\odot$). For ALPs with attractive interactions ($\Lambda < 0$), only a limited range of parameters can sustain stable DANSs, and there will be a maximum $m$ allowed ($m = 5.58 \times 10^{-10}$ eV for $M_{DM} = 0.1$ $M_\odot$). This is expected, because only the repulsive quantum pressure supports the ALP DM from collapsing, while both gravity and self-interaction enhance the collapse. A large enough attractive self-interaction will overcome the repulsive quantum pressure, and the star collapses \citep{Chavanis:2011zm}. It has been shown with dynamical simulations that boson stars with attractive self-interaction will eventually collapse into black holes if they accrete mass over time \citep{Chen:2020cef}. These regions of unstable DANS will be used to constrain the parameter space of bosonic DANS as the 2.6 $M_\odot$ GW190814 secondary.

Now, we focus on a particular case of DANSs in the non-interacting limit with 0.1 $M_\odot$ of DM admixed. We plot the NM mass-radius diagram for these DANSs in Figure \ref{fig:3}. With high-mass bosons ($m > 1.12 \times 10^{-11}$ eV) admixed, the mass-radius relation appears to arise from a softer NM EOS (with no DM admixed). This shift in the mass-radius relation by the admixture of DM could give rise to observable phenomena and help to save some stiff EOSs that do not seem to match with observations. We will further discuss this in the appendix.

To account for the GW190814 secondary, the maximum total system mass $M_{max}$ should be greater than 2.6 $M_\odot$. For large $m$ in the non-interacting limit, $M_{max}$ is even smaller than that of a pure NS. For example, consider the case with $m = 5.58 \times 10^{-10}$ eV, $\Lambda \sim 0$ and $M_{DM} = $ 0.1 $M_\odot$. We could infer from Figure \ref{fig:3} that $M_{max}$ is 2.06 \(M_\odot\), smaller than $M_{max, NS}$ of a pure NS (2.2 \(M_\odot\)). Therefore, there is an upper bound on $m$ for a fixed $M_{DM}$ and $\Lambda$. With $M_{DM} = $ 0.1 $M_\odot$, an increase in $M_{max}$ requires $m < 1.12 \times 10^{-11}$ eV in the non-interacting limit. In this case, the DM is more spread out such that it exerts a smaller gravitational effect on the NS. In the most extreme case where $R_{DM} >> R_{NS}$, $M_{max}$ is simply the sum of $M_{max, NS}$ with no DM admixed and $M_{DM}$.

The radius of the DANS in GW190814 must also not be too large, otherwise it would already be tidally disrupted by the black hole during the phase where the emitted gravitational-wave signal was
detected. We first use Newtonian analysis to estimate that the typical separation of the two objects in GW190814 decreases from about 730 km (with gravitational-wave frequency $\sim$ 30 Hz) to about 330 km (with gravitational-wave frequency $\sim$ 100 Hz). If the DANS is not tidally disrupted at the separation $d \sim 330$ km, the Newtonian tidal force of the black hole at the DANS surface is smaller than the self gravitational force of the star:
\begin{eqnarray}
{2 G M_{BH} R_{2.6} \over d^3 } &\leq&  { G (2.6 \text{ } M_\odot) \over R_{2.6}^2 } , \\
R_{2.6} &\leq& d \left( {2.6 \text{ } M_\odot \over 2 M_{BH} } \right)^{1/3}    \sim 127 \ {\rm km}  ,
\end{eqnarray}
where $M_{BH} = 23$ $M_\odot$ is the mass of the GW190814 primary black hole, and $R_{2.6}$ is the radius with $2.6$ $M_\odot$ enclosed total mass for a DANS model at $M_{max}$. Since $R_{DM} \propto m^{-2}$ for a pure boson star in the non-interacting limit with fixed DM mass, this will impose a lower bound for $m$ with fixed $\Lambda$ and $M_{DM}$. The existence of the innermost stable circle orbit (ISCO) around a black hole also gives rise to yet another constraint on $R_{2.6}$. As the spin parameter of the primary black hole is constrained to $\leq 0.07$ \citep{Abbott:2020khf} and its mass is also about 10 times greater than that of the secondary object, we estimate the location of the ISCO by using a Schwarzschild spacetime solution which gives $R_{ISCO} = 6 M_{BH} \sim 204$ km. Requiring that the DANS surface should be outside the ISCO, we have $R_{2.6} \leq d - R_{ISCO} \sim 126$ km, which turns out to be about the same (coincidentally) as that imposed by the tidal disruption.

\begin{figure}
\centering
\includegraphics[width=1.0\textwidth]{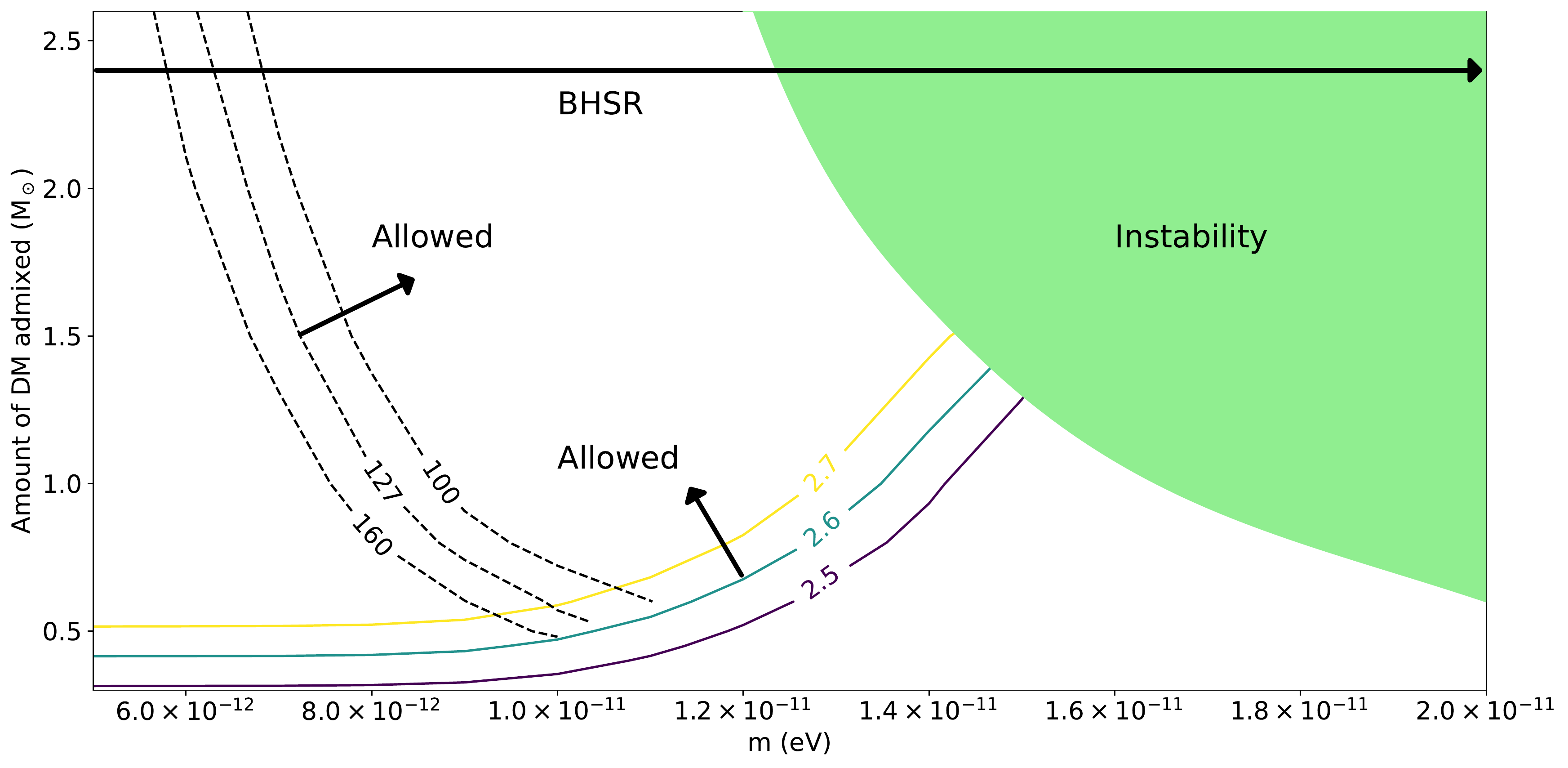}
\caption{Constraints on the DM parameter space for QCD axion DANS to be a candidate for the 2.6 $M_\odot$ GW190814 secondary. The solid lines are contours of different maximum total masses $M_{max} = 2.5$ $M_\odot$, $2.6$ $M_\odot$, and $2.7$ $M_\odot$. The dashed lines are contours of $R_{2.6} = $ 100 km, 127 km, and 160 km. 
The green area marks the parameter space where the bosonic DM will collapse into a black hole due to instability ($M_{max, DM} < M_{DM}$), and the top horizontal arrow marks the range of QCD axion mass $m$ that is ruled out by the constraint imposed by black hole superradiance (BHSR).}
\label{fig:4}
\end{figure}

With the three constraints, we can probe the allowed parameter space for DANS as the GW190814 secondary. For QCD axion, the only two DM parameters are $m$ and $M_{DM}$. In Figure \ref{fig:4}, we show the three constraints discussed above. The solid (dotted) contour lines mark the $M_{max}$ ($R_{2.6}$) achievable for various values of $m$ and $M_{DM}$, and the green region marks the parameters where the bosonic DM will collapse into a black hole due to instability. 
Note that the first constraint in mass ($M_{max} > 2.6$ $M_\odot$) imposes an upper limit in $m$ for fixed $M_{DM}$, and the allowed parameter space is above the $M_{max} = 2.6$ $M_\odot$
contour line, as indicated by the arrow on that line. 
On the other hand, the constraint in radius ($R_{2.6} < 127$ km) imposes a lower limit in $m$ for fixed $M_{DM}$, and the allowed parameter space is on the right of the 
$R_{2.6} = 127$ km contour line, as also indicated by the arrow on that line. 
For the admixture of around 1.0 $M_\odot$ of DM in the DANS, the QCD axion mass is constrained to a narrow range of $8 \times 10^{-12}$ eV $\lesssim m \lesssim 1.3 \times 10^{-11}$ eV. For a larger $M_{DM}$, the constraint on $m$ is loosened. For example, with 1.5 $M_\odot$ of DM admixed, $7 \times 10^{-12}$ eV $\lesssim m \lesssim 1.4 \times 10^{-11}$ eV is allowed.

As we discussed in the introduction, there is another constraint on $m$ due to the fact that the existence of light bosonic DM will cause a black hole to spin-down due to the effect of black hole superradiance \citep{Arvanitaki:2014wva}. The non-observation of this spin down effect in stellar black holes then leads to an exclusion mass range of $6 \times 10^{-13}$ eV $\leq m \leq 2 \times 10^{-11}$ eV \citep{Arvanitaki:2014wva}, which is marked by the horizontal arrow at the top of Figure \ref{fig:4}. It can be seen that all the parameter values of $m$ and $M_{DM}$ for the QCD axion DANS to be a candidate for the 2.6 $M_\odot$ compact object are excluded.

\begin{figure}
\centering
\includegraphics[width=1.0\textwidth]{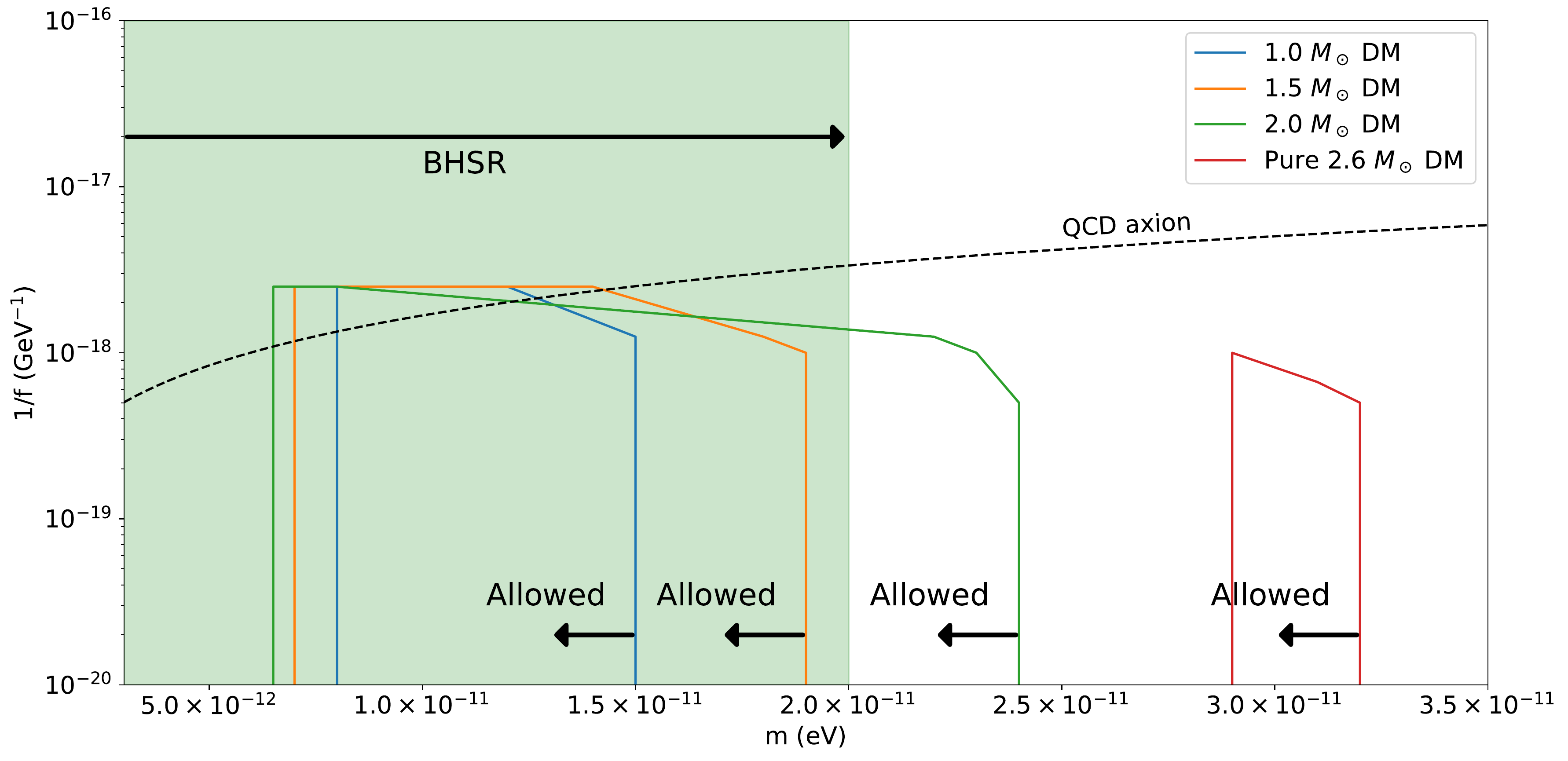}
\caption{Constraints on the DM parameter space for ALP DANS to be a candidate for the 2.6 $M_\odot$ GW190814 secondary with different total DM masses. 
The colour lines enclose the allowed parameter spaces for different
$M_{DM}$. The QCD axion relation (Eq.~(\ref{eqn:3})) between $m$ and $f$ is represented by the 
dashed line. 
The green region marks the range of ALP mass $m$ and decay constant $f$ that is ruled out by black hole superradiance (BHSR).}
\label{fig:5}
\end{figure}

For ALPs, the decay constant $f$ is an extra parameter. For each pair of $m$ and $f$, we model a DANS with $M_{DM} = 1.0, 1.5,$ and $2.0$ $M_\odot$. We then use the three constraints in mass, radius, and stability of the DANS to get the allowed parameter space in $m$, $f$, and $M_{DM}$, shown in Figure \ref{fig:5}. The green region in the figure is excluded by the black hole superradiance.
The different colour lines enclose the allowed parameter spaces
for different $M_{DM}$. For instance, the region enclosed by the blue line represents the 
allowed parameter space for $M_{DM}=1.0$ $M_\odot$.
It is clear that for $M_{DM} \leq 1.5$ $M_\odot$, all values of $m$ and $f$ are excluded by the black hole superradiance. However, for $M_{DM} = 2.0$ $M_\odot$, a narrow window of $2 \times 10^{-11}$ eV $\leq m \leq 2.4 \times 10^{-11}$ eV and $f \geq 8 \times 10^{17}$ GeV is allowed. 
The realtion between $m$ and $f$ (Eq. (\ref{eqn:3})) for QCD axions is indicated by the dotted line. One can see that even with an arbitary amount of admixed DM, the QCD axion DANS is ruled out, which is the same conclusion as in Figure \ref{fig:4}. 

In addition, we consider the case of a pure ALP DM star without NM, for which the allowed parameter space is enclosed by the red line in Figure \ref{fig:5}.
One can see that a pure QCD axion DM star also cannot be a candidate for the 2.6 $M_\odot$ compact object, as the dotted line indicating the QCD axion $m$-$f$ relationship does not intersect with the allowed region marked by the red line. However, the ALP DM star with $f \geq 8 \times 10^{17}$ GeV is not ruled out by the black hole superradiance in the mass range $2.9 \times 10^{-11}$ eV $\leq m \leq 3.2 \times 10^{-11}$ eV.

\begin{figure}
\centering
\includegraphics[width=1.0\textwidth]{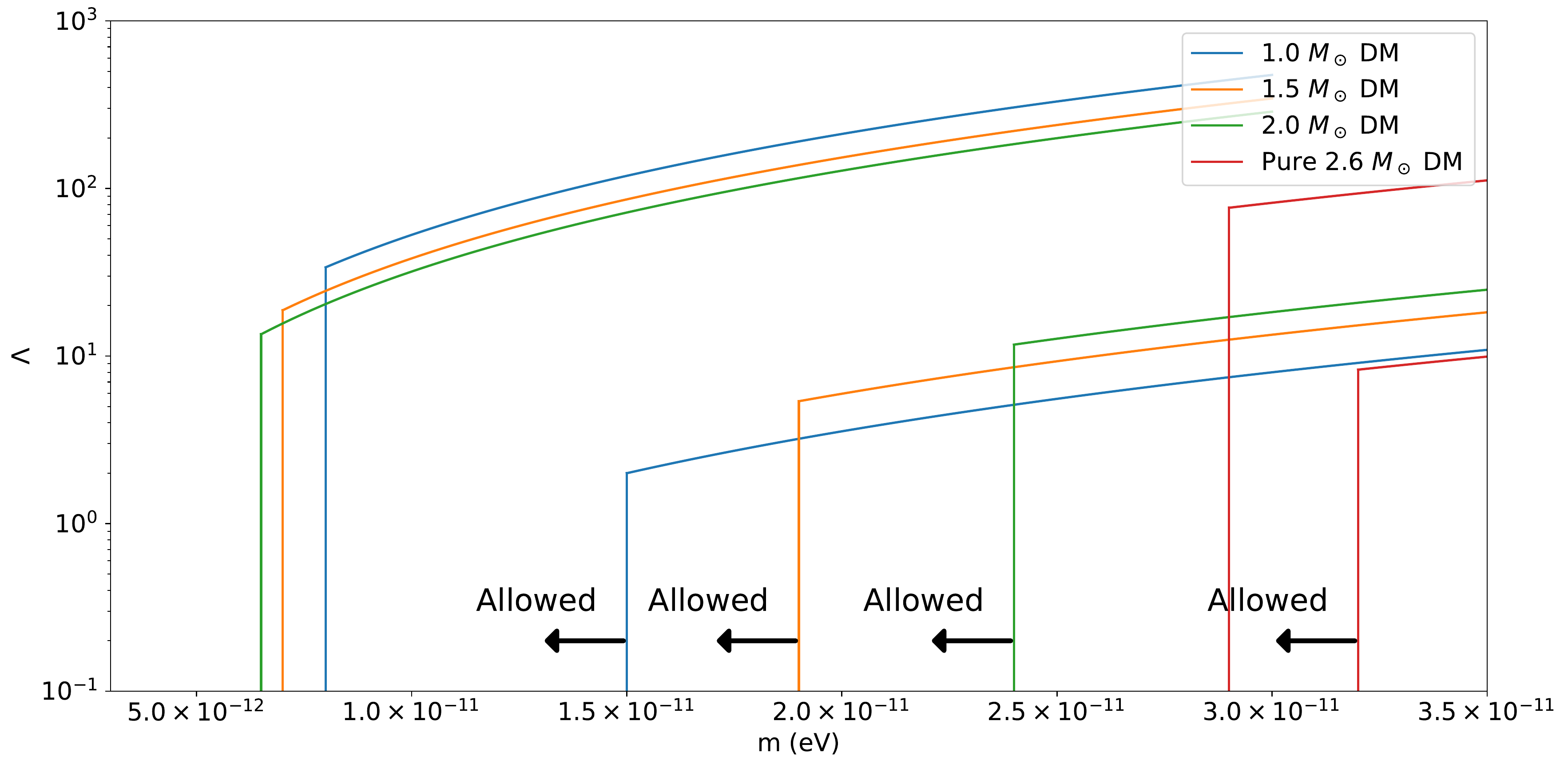}
\caption{Constraints on the DM parameter space for repulsive self-interacting bosonic DANS to be a candidate for the 2.6 $M_\odot$ GW190814 secondary with different total DM masses. The colour lines enclose the allowed parameter spaces for different
$M_{DM}$.}
\label{fig:8}
\end{figure}

The same three constraints in mass, radius, and stability discussed above can be applied to repulsive self-interacting bosons. Figure \ref{fig:8} shows the constraint plot similar to Figure \ref{fig:5}, where different colour lines enclose the allowed parameter space in $m$, $\Lambda$, and $M_{DM}$ for the 2.6 $M_\odot$ compact object in GW190814 to be a repulsive self-interacting bosonic DANS or pure repulsive self-interacting boson star. However, unlike the case for ALPs, $m$ is only constrained from below. With $M_{DM} = 0.1$ $M_\odot$, $m \geq 8 \times 10^{-12}$ eV, and for the case of a pure repulsive self-interacting boson star, $m \geq 2.9 \times 10^{-11}$ eV. It is known that in the TF-limit, the only parameter that determines the properties of a DANS is $m/\sqrt{\Lambda}$ for a given $M_{DM}$. As a result, a large $m$ is always possible given a large $\Lambda$. Nonetheless, we can constrain the value $m/\sqrt{\Lambda}$ in the TF-limit for different $M_{DM}$. For example, with $M_{DM} = 1.0$ $M_\odot$ in the TF-limit, we have $1.38 \times 10^{-12}$ eV $\leq m/\sqrt{\Lambda} \leq 1.06 \times 10^{-11}$ eV, and with a pure repulsive self-interacting boson star, we have $3.31 \times 10^{-12}$ eV $\leq m/\sqrt{\Lambda} \leq 1.11 \times 10^{-11}$ eV.

\section{Discussion} \label{sec:discussion}

If the 2.6 $M_\odot$ GW190814 secondary is indeed a DANS, a large amount of DM is needed to be 
admixed to NM. Various studies have shown that bosonic DM, regardless of its self-interaction, can be accreted onto a fermion star (e.g., \citep{DiGiovanni:2020frc}) to form a stable fermion-boson star
(at least in principle). Furthermore, a NS can produce and trap scalar bosons through the process of superradiance similar to a black hole \citep{Day:2019bbh}. However, the amount of DM that can be accumulated in a compact star depends on the location of the star in the DM halo and its evolution history \citep{Ivanytskyi:2019wxd}. There is also the possibility that the GW190814 secondary can be a pure ALP star as suggested in Figure~\ref{fig:5}. If this is the case, it will be as interesting as DANS because it will be the first exotic boson star ever discovered.

The relevant range of $m$ that can form stellar-mass DANSs is much lower than the experimental search range for axions, from $\mu$eV to meV. While a common technique for the direct search of axions is through detecting the photons from the axion-photon conversion, a rough estimation on the axion-photon coupling constant $g_{a\gamma \gamma}$ using $m = 2 \times 10^{-11}$ eV gives $g_{a\gamma \gamma} \sim 2.8 \times 10^{-21}$ GeV$^{-1}$, much below the sensitivity of current experiments, such as the ADMX experiment that can reach $g_{a\gamma \gamma} > 10^{-15}$ GeV$^{-1}$ in the 2.81 $-$ 3.31 $\mu$eV range \citep{Braine:2019fqb}. Our interested range of $m$ implies converted photons at $\sim$ 4800 Hz in the very-low radio wave frequency range, which is not observable by ground telescopes. If more compact objects with mass $> 2.2$ $M_\odot$ are found, it may open up a new search window for axions in the peV $-$ neV range.

Even if no DANS is found though observations, such as looking for the altered supernova light curve due to the decrease in $M_{max, WD}$ by the admixture of DM as we discuss in the appendix, it could imply a large $R_{DM}$ that hardly affects the NM properties. For example, with $M_{DM} = $ 0.1 $M_\odot$, $m$ ($m / \sqrt{\Lambda}$) is constrained to be smaller than \(1.12 \times 10^{-11}\) eV ($1.12 \times 10^{-13}$ eV) in the non-interacting (TF) limit as inferred from Figure \ref{fig:1}. In this case, models such as the ultra-light bosons \citep{Ji:1994xh} are favoured.

We emphasise that different DM models, such as self-interacting bosons, mirror particles \citep{Sandin:2008db}, and ideal fermi gas \citep{Leung:2012vea, Leung:2013pra}, yield similar results for DANSs or DAWDs. In all the above models, \(M_{max,NS}\) or \(M_{max,WD}\) decreases and the visible radius of the compact object shrinks compared to the case without DM admixed. This is not surprising, as the NM only feels the presence of DM through gravity, so that only the density distribution of the DM matters. As a consequence, it is not easy to select DM models by observing DM-admixed stars. Still, given a DM model, one can use the same constraints on mass, radius and stability of the DM-admixed star to constrain properties of DM including particle mass and self-coupling constant, both for bosonic and fermionic DM.

\section{Conclusion} \label{sec:conclusion}

In summary, by combining the constraints in mass, radius, stability of a 2.6 $M_\odot$ DANS with that on $m$ by black hole superradiance \citep{Arvanitaki:2014wva}, we find that the 2.6 $M_\odot$ GW190814 secondary cannot be a QCD axion DM admixed compact star. For ALPs, a small window of mass ($2.0 \times 10^{-11}$ eV $\leq m \leq 2.4 \times 10^{-11}$ eV) is allowed when at least 2.0 $M_\odot$ of DM with decay constant $f \geq 8 \times 10^{17}$ GeV is admixed with NM. A pure ALP star is allowed in the mass range $2.9 \times 10^{-11}$ eV $\leq m \leq 3.2 \times 10^{-11}$ eV with $f \geq 8 \times 10^{17}$ GeV.

\appendix

\section{How DANSs could help bypass constraints on NM EOS} \label{sec:appA}

If DM is admixed, the constraint of the NS mass-radius relation determined by the NM EOS could be bypassed, and EOSs such as APR which do not satisfy all observational results may be saved. The mass-radius relation of DANSs with different amounts of DM admixed will become a band instead of a line, which, if observed, could also be a hint of the existence of DANSs \citep{Ciarcelluti:2010ji}. Since independent WD mass and radius observations using eclipsing binaries have become more sophisticated \citep{Parsons}, DM-admixed WDs (DAWDs) may also be discovered if they exist.

\begin{figure}
\centering
\includegraphics[width=0.8\textwidth]{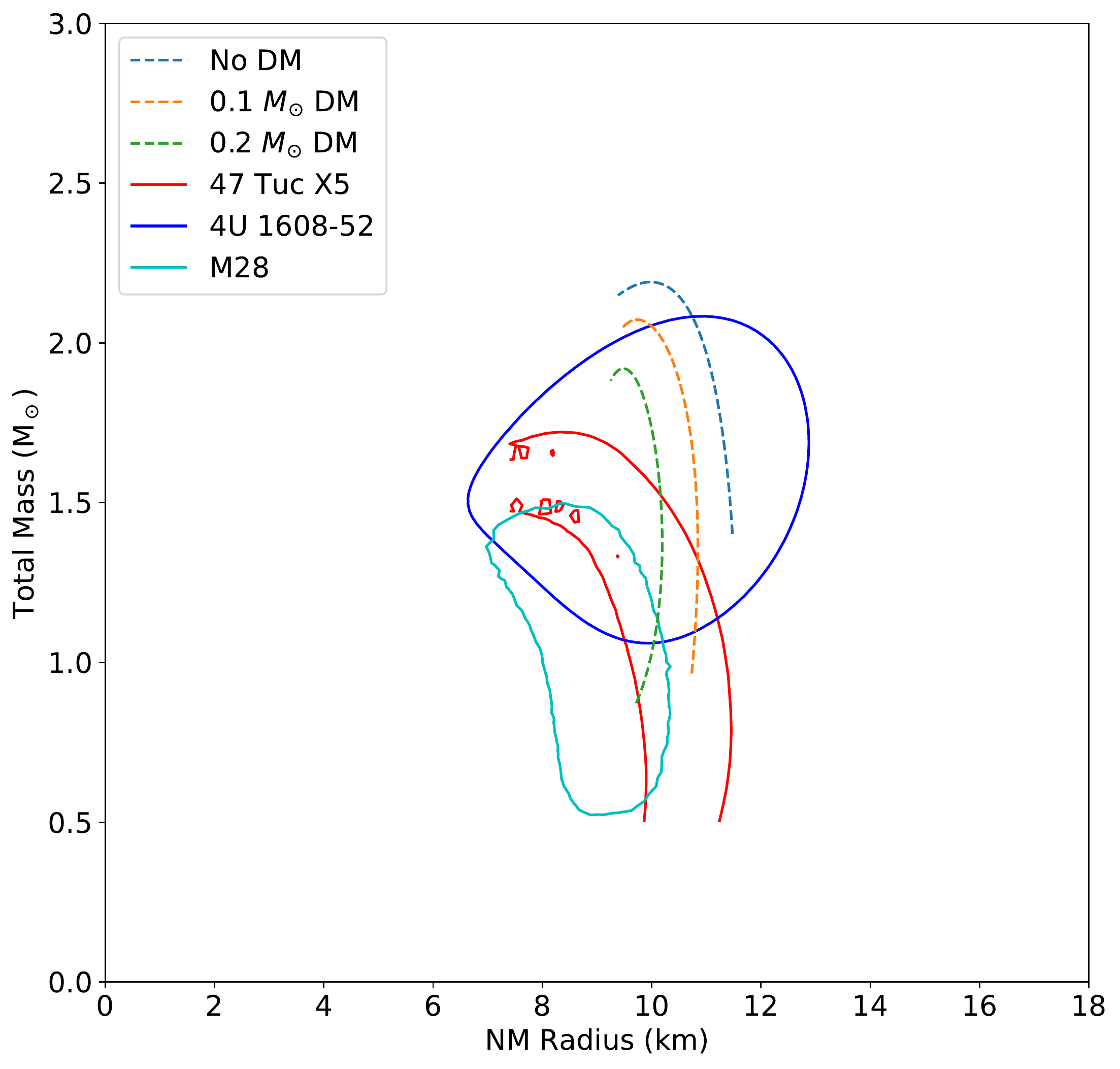}
\caption{Mass-radius relations of DANSs with different amounts of DM admixed, using the APR EOS for NM. $m = 10^{-8}$ eV and $\Lambda = 1.47 \times 10^5$. Observational data for 3 NSs are shown for comparison, with the contours showing the boundaries of 1-\(\sigma\) significance.}
\label{fig:10}
\end{figure}

We illustrate such a scenario by choosing a high boson mass \(m=10^{-8}\) eV with large self-coupling constant \(\Lambda= 1.47 \times 10^{5}\) in the TF limit. The mass-radius relation for DANSs with different amounts of DM admixed is shown in Figure \ref{fig:10}, assuming the APR EOS for the NM. In the same figure, we also plot three NS observational data \citep{Ozel:2015fia}, for the highest mass 4U 1608-52, intermediate mass 47 Tuc X5 and smallest mass M28. We can see that with no DM admixed, the mass-radius relation is only consistent with the 4U 1608-52 observation. However, with 0.1 \(M_\odot\) of DM admixed, the line shifts to the bottom-left direction, matching with both 47 Tuc X5 and 4U 1608-52 observations. Eventually, with 0.2 \(M_\odot\) of DM admixed, all three NS observations can be accounted for with the same APR EOS.

\begin{figure}
\centering
\includegraphics[width=1.0\textwidth]{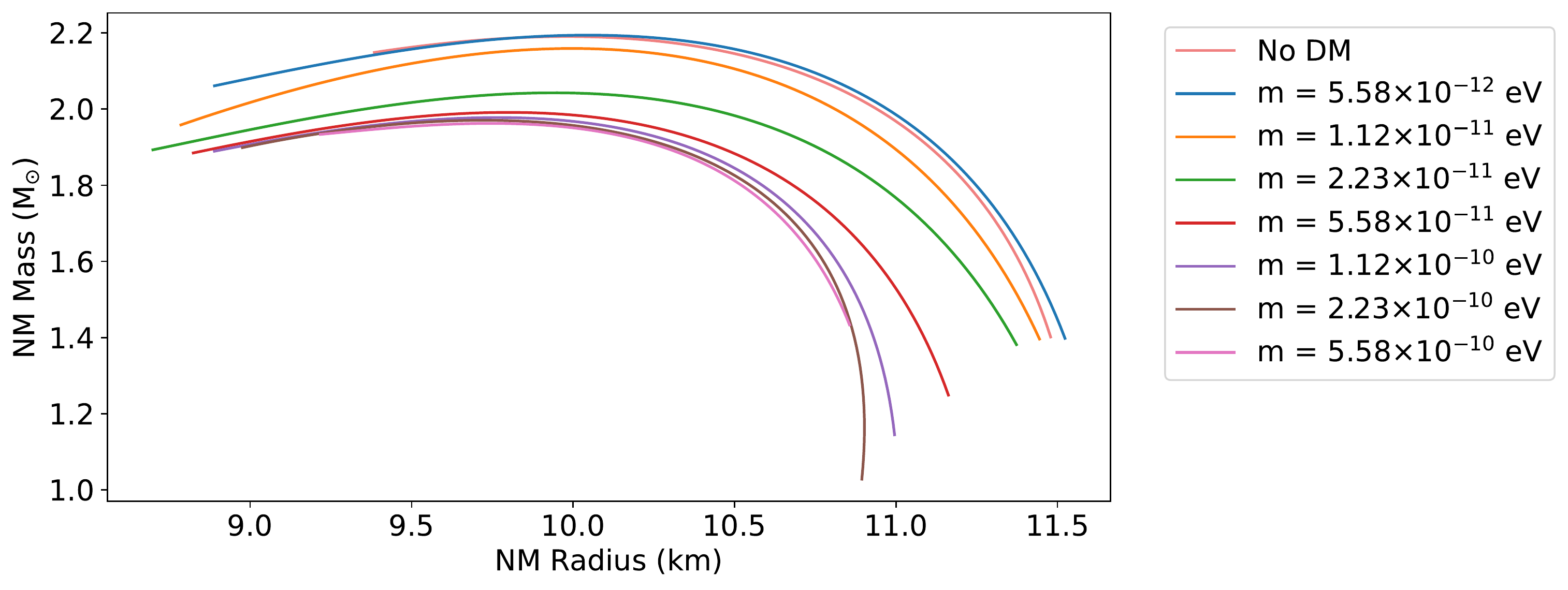}
\caption{NM mass-radius relation of DANSs with different non-interacting bosonic DM particle masses to make up a total of 0.1 \(M_\odot\) DM mass.}
\label{fig:3}
\end{figure}

\begin{figure}
\centering
\includegraphics[width=0.9\textwidth]{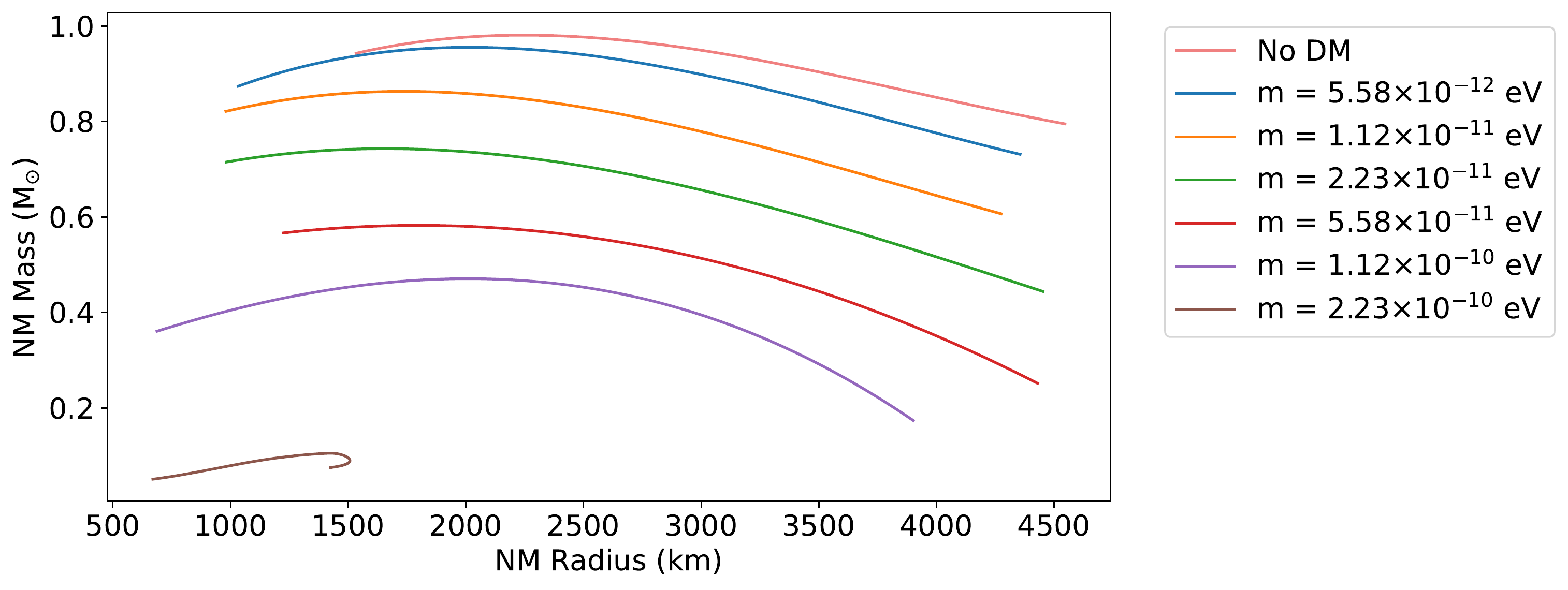}
\caption{Same as Figure \ref{fig:3}, but for DAWDs.}
\label{fig:7}
\end{figure}

However, the maximum WD mass $M_{max,WD}$ will be significantly decreased if the same amount of DM is admixed to a WD. For example, from Figure \ref{fig:3} one can see that admixing 0.1 \(M_\odot\) of \(m = 1.12 \times 10^{-10}\) eV DM to an NS would reduce \(M_{max,NS}\) to 1.99 \(M_\odot\), making the NM EOS looks softer. However, the same amount of DM admixed to a WD will reduce \(M_{max,WD}\) to 0.46 \(M_\odot\), as inferred from Figure \ref{fig:7}, and the DAWD can easily become a Type Ia supernova. Even though one cannot measure \(M_{max,WD}\) directly, it is highly related to Type Ia supernovae, and there are already a few studies of DM-admixed supernovae \citep{Leung:2019ctw, Chan:2020ijt}. Since admixing DM could reduce \(M_{max,WD}\), which changes the initial conditions of the Type Ia supernovae, the resulting supernova light curve is dimmer than that with no DM admixed \citep{Chan:2020ijt}.

\acknowledgments

This work is partially supported by grants from the Research Grants Council of the Hong Kong Special Administrative Region, China (Project No.s AoE/P-404/18, 14300320).


\bibliography{sample63}{}
\bibliographystyle{aasjournal}



\end{document}